\documentclass[aps,preprint,groupeaddress,showpacs]{revtex4}

\usepackage{amsmath} 
\usepackage{amssymb} 
 \usepackage{amsfonts}
\usepackage[dvips]{graphicx} 
\usepackage[]{epsfig} 

\begin{document}

\title{Ionic force field optimization based on single-ion and ion-pair  solvation properties}
\author{Maria Fyta$^1$, Immanuel Kalcher$^1$, Joachim Dzubiella$^1$, Lubo$\check{s}$ Vrbka$^2$, and Roland R. Netz$^1$}

\affiliation{
$^1$Physics Department (T37), Technical University of Munich, 85748 Garching, Germany \\
$^2$Institute for Physical and Theoretical Chemistry, University of Regensburg, 93040 Regensburg, Germany
}

\begin{abstract}
Molecular dynamics simulations of ionic solutions depend sensitively on the force fields employed for the ions. 
To resolve the fine differences between ions of the same valence and roughly similar size and in 
particular to correctly describe ion-specific effects, it is clear that accurate force fields are necessary.
In the past, optimization strategies for  ionic force fields either considered single-ion properties
(such as the solvation free energy at infinite dilution or the ion-water structure) or ion-pair properties (in the form 
of ion-ion distribution functions).  In this paper we investigate strategies to optimize ionic force fields
based on single-ion  and ion-pair properties simultaneously. To that end,
we simulate five different  salt solutions, namely CsCl, KCl, NaI, KF, and CsI, 
at finite ion concentration. The force fields of these ions are systematically varied under the constraint
that the single-ion solvation free energy matches the experimental value, which reduces the two-dimensional
$\{\sigma,\varepsilon\}$ parameter space of the Lennard Jones interaction to a one dimensional line for each ion.
From the finite-concentration simulations, the pair-potential is extracted and the osmotic coefficient is calculated,
which is compared to experimental data. We find a strong dependence of the osmotic 
coefficient on the force field, which is remarkable as the single-ion solvation free energy and the ion-water
structure remain invariant under the
parameter variation. Optimization of the force field is achieved for the cations Cs$^+$ and K$^+$, while
for the anions I$^-$ and F$^-$ the experimental osmotic coefficient cannot be reached. This suggests that 
in the long run,
additional parameters might have to be introduced  into the modeling, for example by modified mixing rules.
\end{abstract}

\maketitle

\section{Introduction}

Aqueous electrolyte solutions are of fundamental importance not only in physical chemistry 
but also for biological function and technological applications. In biology, the presence of ions, 
specifically of the monovalent ions K$^+$, Na$^+$, Cl$^-$, has significant effects on the stability, structure, and 
function of nucleic acids and proteins and the regulation of biomolecular processes~\cite{anderson95,baldwin:biophys:96,alberts}. In technological applications, ions can play an important role in chemical reactions by influencing their rates, 
as well as in  controlling the solubility of various co-solutes~\cite{kunz,kumar01}. For salt concentrations larger than $\sim$~10 mM salt effects are typically  ion-specific even for simple bulk properties such as  the osmotic pressure, which, in turn, can be highly relevant for transport and function in biomolecular  systems~\cite{anderson95,alberts}. The molecular understanding and prediction of the complex and often context-dependent effects of aqueous electrolytes poses a challenging task to the scientific and, in particular, theoretical 
community~\cite{collins2,jungwirth}.

The successful molecular modeling of ionic effects  typically involves computer 
simulations in which the ionic and water degrees of freedom are explicitly resolved and evolved by 
a set of effective, classical interactions: the simulation  {\it force field}.  
Quite commonly used force fields are pair-wise additive and non-polarizable to keep the parameter space small. 
On that level, the atoms are characterized by  (partial) Coulombic point charges $q_{i}$, 
excluded-volume radii, and dispersion attraction strengths. 
In standard protocols, the non-electrostatic interaction for atoms $i$ and $j$ at a distance $r_{ij}$ is modeled by a 
pairwise Lennard-Jones (LJ) interaction of the form 
\begin{eqnarray}
V_{\rm LJ}(r_{ij})=4\varepsilon_{ij}\left[\left(\frac{\sigma_{ij}}{r_{ij}}\right)^{12}-\left(\frac{\sigma_{ij}}{r_{ij}}\right)^{6}\right],
\label{eq:lj}
\end{eqnarray}
with two free parameters, the interaction length $\sigma_{ij}$ and  the energy scale $\varepsilon_{ij}$, per pair of atoms. 
The whole set  $\{\sigma_{ij},\varepsilon_{ij},q_{i}\}$ with $i,~j=1,...,M$,  defines the total force field behind the 
nonbonded inter- and intra-molecular molecular dynamics (MD) interactions for M atomic species. 
Typically, the vast number of parameters is reduced by using 
heuristic  mixing rules for the cross interactions ($i\neq j$) so that the only remaining parameters are
the diagonal coefficients $\sigma_{ii}$ and $\varepsilon_{ii}$.
The common  mixing rules are $\varepsilon_{ij}=\sqrt{\varepsilon_{ii}\varepsilon_{jj}}$, 
and either $\sigma_{ij}=(\sigma_{ii}+\sigma_{jj})/2$, constituting the Lorentz-Berthelot (LB) mixing rules, 
or  $\sigma_{ij}=\sqrt{\sigma_{ii}\sigma_{jj}}$, defining the geometric mixing rules \cite{frenkelsmit}. 
It is assumed that the force fields take implicitly into account the polarizability, as well as many-body effects. 
In fact, potentials which account for electric polarizability do not strictly seem to be required for modeling ion pairing:
It has been shown, that for mono- and divalent ions even the first hydration shell is not significantly polarized compared to water molecules in the bulk \cite{krekeler07}, though it should be kept in mind that a force field with more parameters
has the principal possibility to be more accurate. 
For water usually simple point charge models (e.g., TIP3P or SPC/E) are used in which oxygen and hydrogen atoms are resolved~\cite{berendsen:jpc,tip3p}. The latter are connected by rigid intramolecular bonds and carry  partial charges optimized in such a way that a few important water properties (density, structure, surface tension, dielectric constant) are well reproduced.  

Throughout the years, there have been numerous attempts to systematically optimize ionic force fields. 
In principle, to properly describe the interactions between ions and between  ions and water, 
a high level of quantum theory is required, which however turns out to be computationally too demanding for many-ion systems. 
Because of this, but also due to the weak computer power in the early days of force field development, it has become
 a common habit to parametrize the empirical force fields of ions based 
 on single ion properties, such as ion solvation free energies or hydration structure in small water clusters, see for example Ref. \cite{dang}. Force fields based on this procedure, though, often fail to reproduce realistically the ion-ion fluid structure and thermodynamics of the electrolytes at non vanishing concentrations, even for simple ionic solutions such as 
NaCl \cite{lyubartsev:pre:97,hess:jcp:06}.  As has been recognized in the literature, there is a strong sensitivity of solution thermodynamics~\cite{friedman:71,pettitt:jcp:86,dang:jcp:92} and contact ion pairing~\cite{collins1,dill09} to small changes in the effective pair potential between the interacting ions. Ion force field development is, thus, 
a difficult task and remains an active field of research~\cite{ponder03,patra:jcc:04}. 
Recently, two studies have revisited and systematically explored single ion hydration free energies  by 
scanning through a wide region of the  LJ parameter space $\{\sigma,\varepsilon\}$~\cite{suk:jpcb,horinek09}. 
This was triggered by the observation that while traditionally force field parameters have been adjusted in order to correctly 
reproduce single-ion free energies of solvation, one experimental observable is clearly
not sufficient to unequivocally fix the two parameters $\varepsilon$ and $\sigma$. In \cite{suk:jpcb}, crystal lattice energies have been used as a second independent optimization target. In \cite{horinek09}, the single-ion solvation entropy and the effective solution ion size (as constructed
from the ion-water radial distribution function) have been used, though the simultaneous optimization of two parameters, especially for the cations was problematic.
It was seen that the three observables considered, namely the free energy of solvation, the entropy of solvation, and
the effective ion size, roughly matched the experimental values on a whole curve in the $\{\sigma,\varepsilon\}$ parameter plane,
not allowing to single out one of the cationic parameter combinations as truly optimal \cite{horinek09}.
It would be desirable to nail down the final cationic parameter set by benchmarking to 
collective, thermodynamic solution properties such as the electrolyte 
activity or osmotic pressure. Weerasinghe and Smith introduced and carried out this idea for NaCl solutions by  reproducing Kirkwood-Buff (KB) integrals as determined by experiments, ensuring that a good representation of solution activity is obtained~\cite{smith:jcp:03}.  The same approach has been used recently to investigate the cation-specific binding with protein surface charges~\cite{hess09}.
However, the parametrization, which involves a free fit parameter in the mixing rule, 
does not conserve the free energy of solvation of single ions. It is therefore an interesting question, whether one can 
{\em simultaneously} describe single ion properties (such as the free energy of solvation) and ion-pairing properties
by choosing optimized parameters for $\{\sigma,\varepsilon\}$ alone or whether an additional parameter has to be
introduced, e.g. in the form  of a generalized mixing rule as was recently suggested~\cite{hess09}.

An alternative method for benchmarking  MD force fields
has been introduced  by Hess {\it et al.}~\cite{hess:prl:06} and Kalcher and Dzubiella ~\cite{kalcher09}. Here,
effective, MD-derived ion-ion pair potentials are used in a many-body corrected virial route to obtain osmotic pressures. 
The electrolyte structure at a given concentration, which forms an input to the virial equation, can 
be obtained directly from an MD simulation or approximately, but with much less computational 
effort, from simulations with implicit solvent~\cite{hess:prl:06,hess:jcp:06} or hypernetted-chain (HNC) integral equation theory~\cite{vrbka09}. It was shown that the KB derived NaCl force field~\cite{smith:jcp:03} 
and a few alkali-Cl force fields  proposed by Dang~\cite{dang:jacs:95} could reproduce experimental 
osmotic coeffcients in SPC/E water quite well, while  others badly failed~\cite{hess:prl:06,kalcher09}. 
The reason for the failure of some of the force fields when considering ion-pairing properties did not become clear.

In this work, we explore the optimization of ionic force fields based on single-ion and ion-pair
thermodynamic properties, using the infinite-dilution solvation free energy and the 
finite-concentration electrolyte osmotic pressure as benchmarks.
As for NaCl reasonable force fields exist, we use for Na$^+$ and Cl$^-$  the parameters
given by Dang \cite{dang:jacs:95,dang:92}, 
and focus on the salts  KCl, NaI, KF, CsCl, and CsI in SPC/E water
and systematically vary the force fields of K$^+$, Cs$^+$, F$^-$, and I$^-$.
To satisfy the  experimental ion solvation free energies,
we confine our search in LJ parameter space to the experimental equi-solvation free energy lines in 
$ \varepsilon-\sigma$ space as calculated  previously~\cite{horinek09}.
In this paper we do not vary systematically the mixing rule  
and in most simulations use the LB mixing rules. 
We apply the procedure proposed by 
Kalcher and Dzubiella~\cite{kalcher09} to calculate the electrolyte osmotic coefficients at a finite concentration of 1 M for 
a wide range of LJ parameters and compare to experiments. Our calculations are accompanied 
by HNC integral equation calculations that allow to efficiently cover and investigate a wide range  
of electrolyte concentrations.
We systematically explore (a) whether ionic force field optimization is possible consistently  for both single ion solvation energies and  collective electrolyte thermodynamics without loosening the parameter space constraint  given by the mixing rule, and (b) how the thermodynamics of electrolyte solutions react to a  change of the LJ parameters along the equi-solvation free energy path. 
As a main result, this simultaneous optimization of the force field seems possible  for the cations Cs$^+$ and $K^+$, while
for the anions I$^-$ and F$^-$ the results are less promising. This suggests that in the long run, and in
order to consistently describe finite-concentration electrolyte thermodynamics, 
 the mixing rules have to be modified and systematically optimized. By calculating osmotic coefficients for 
a solution of CsI we  also investigate the transferability 
 of ion parameters for Cs$^+$ and I$^-$ that have been separately optimized by matching osmotic coefficients
 for CsCl and NaI. Modulo the previously mentioned restricting comments on the optimization of I$^-$ parameters,
 ion parameters seem transferable in the sense that trends in the osmotic coefficients of certain ions
 also are found when those ions are assembled into different  ion pairings.

\section{Methods\label{sec:method}}

\subsection{Simulation details}

We perform atomistic simulations using the Molecular Dynamics (MD) package 
GROMACS \cite{gromacs1,gromacs2} in the ($N,p,T$) canonical ensemble, 
for which the particle number $N$, as well as the pressure $p=1$~bar and temperature $T=300K$ are held constant 
using a Berendsen barostat and thermostat \cite{berendsen:jcp}. The simulation box is cubic, with an edge length of  $L \simeq 4$ nm, and periodically repeated in all three dimension and
includes explicit ions and  a total number of about 2000 SPC/E  water molecules \cite{berendsen:jpc}. 
Finite size effects are not significant for these sizes, as shown by a previous study on similar systems and sizes \cite{kalcher09}. 
The three dimensional particle-mesh Ewald sum is used for the electrostatics \cite{essmann:jcp} with
 a grid spacing in Fourier space of $0.12$ nm in all three directions. We use an interpolation order of 4, a distance cutoff of $0.9$ nm for the real-space interactions, and a relative strength of the electrostatic interaction at the cutoff of $10^{-5}$. Typical times for the simulations for gathering statistics are 150 ns for low concentrations and 40-50 ns for moderate concentrations.

Five different salt solutions were simulated, CsCl, KCl, NaI, KF, and CsI 
at densities of 0.3~M and 1~M with 12 and 39 pairs of ions in the solution, respectively. Here, the concentration M denotes molarity (mol/l). For the ions, charged and non-polarizable spheres were used interacting with the LJ potential (Eq. \ref{eq:lj}). For the SPC/E water the LJ parameters are $\varepsilon_{OO}=0.6500$ (kJ/mol) and $\sigma_{OO} =0.3166$ (nm) and are assigned only to the oxygen molecule of water (no parameters are related to the two hydrogen atoms). Point charges of $q_O=-0.8476e$ and $q_H=+0.4238e$ are assigned to the oxygen and hydrogen atoms, respectively. For the ions we varied both parameters $\varepsilon_{iO}$ and $\sigma_{iO}$ and present the analysis in Section \ref{sec:result}. Only for sodium (Na$^{+}$) and chloride (Cl$^{-}$) we used fixed parameters, those given by Dang \cite{dang:jacs:95,dang:92}, as they have been proven to be consistent in determining thermodynamic properties \cite{kalcher09,hess09} and give reasonable hydration energies~\cite{horinek09,suk:jpcb}. The LJ parameters are for Na$^+$, $\varepsilon_{Na,O}=0.5216$ (kJ/mol) and $\sigma_{Na,O}=0.2876$ (nm) and for Cl$^-$, $\varepsilon_{Cl,O}=0.5216$ (kJ/mol) and $\sigma_{Cl,O}=0.3785$ (nm). 
In this notation, the subscripts $iO$ denote the parameters between ion {\it i} and the oxygen atom of the SPC/E water model. For the cross interactions between two ions  we use the Lorentz-Berthelot mixing rules (except where noted otherwise).

For NaCl we also tried  a different  force field that was optimized based on Kirkwood-Buff integrals \cite{smith:jcp:03}. The parameters of this force field are for Na$^+$, $\varepsilon_{Na,O}=0.342$ (kJ/mol) and $\sigma_{Na,O}=0.279$ (nm) and for Cl$^-$, $\varepsilon_{Cl,O}=0.547$ (kJ/mol) and $\sigma_{Cl,O}=0.373$ (nm). That force field involves a modified mixing rule for the relation between ion-ion and ion-water interaction, as it was not possible to fit the experimental data without breaking the geometric mixing rule~\cite{smith:jcp:03}.

\subsection{Effective ionic pair potentials}
\label{pp}
We begin with a brief description of the derivation of the effective ionic (infinite dilution) pair potentials for the salts studied here,
for more details see \cite{kalcher09}. The pair potentials are derived from  
the radial distribution functions (rdfs) obtained within finite-concentration MD simulations. 
The rdf between a pair of atoms or ions $i$ and $j$ at distance $r$ is defined as $g_{ij}(r;\rho)$ at a given salt concentration $\rho$. The potential of mean force (pmf) $w_{ij}(r;\rho)$ at  concentration $\rho$ results from the rdf through a Boltzmann inversion \cite{mayer,mcmillan}:
 \begin{equation}
\beta w_{ij}(r;\rho)=-\ln[g_{ij}(r;\rho)]=\beta \big[w_{ij}^{sr}(r;\rho)+w_{ij}^{lr}(r;\rho)\big],
\label{eq:pmf}
 \end{equation}
where $\beta=1/k_B T$ is the inverse thermal energy.  The pmf can be decomposed into a short-ranged  and a  long-ranged  contribution, $w_{ij}^{sr}(r;\rho)$ and $w_{ij}^{lr}(r;\rho)$, respectively~\cite{gavryushov:jpcb:06,gavryushov:jpcb:06b,HansenMcDonald,perkyns92}. The long-ranged  part of the pmf is a non-specific Debye-H\"uckel potential and can be subtracted from $w_{ij}(r;\rho)$ leading in this way to the short-ranged part of the pair potential as detailed previously~\cite{kalcher09},
\begin{equation}
w_{ij}^{sr}(r;\rho)=w_{ij}(r;\rho)-w_{ij}^{DH}(r;\rho).
\label{eq:vsr}
 \end{equation}
In the low concentration limit, the pmf between two ions reduces  to their effective pair potential and the decomposition described in Eq.~(\ref{eq:pmf}) can be written as:
\begin{equation}
\beta V_{ij}^{\rm eff}(r)=\beta  V_{ij}^{sr}(r)+z_iz_j\lambda_B(0)/r
\label{eq:veff}
 \end{equation}
where the  potential is split into the short-ranged part of the pair potential, $V_{ij}^{sr}(r)$, and the usual Coulombic part. In this equation, $z_i,~z_j$ are the valencies for the two ion types, respectively, and $\lambda_B(\rho)$ is the concentration dependent Bjerrum length with an infinite-dilution (pure water) value of $\lambda_{B}(0)=0.78$~nm for SPC/E water  (about 10\% larger than the real water value) \cite{kusalik94}.  The key assumption of our derivation is now that the short-ranged part of the pair potential, $V_{ij}^{sr}$,
can be extracted from the finite concentration pmf, $V_{ij}^{sr}(r)\simeq w_{ij}^{sr}(r;\rho)$, as calculated in (3), at not too high concentration. This is  a good approximation, as long as the density is smaller than the density
where the hydration layers of ions begin to overlap,
 which has  been found to be betwen 0.5 and 1 M~\cite{kalcher09}. On empirical grounds, the above procedure for the derivation of the ionic pair potentials works well for rdfs generated at a finite concentration of $\rho \simeq 0.3$~M. Here, 
 $V_{ij}^{sr}(r) \simeq w_{ij}^{sr}(r;\rho)$ is well fulfilled and accurate rdfs can be sampled with good statistics~\cite{kalcher09}.  In the following, the total effective pair potential  $V_{ij}^{\rm eff}(r)$ is used as an input to the pressure calculations by the virial route
 and as an input to the HNC method.

\subsection{Virial route to the osmotic coefficient $\phi(\rho)$}

The optimization of the ionic force fields for the salts studied in this work is done by comparing the derived osmotic coefficients to their experimental values. We use the virial route to calculate osmotic coefficients as was done previously~\cite{kalcher09}. The osmotic pressure $\Pi=2\rho \phi(\rho)/\beta$ of the ionic solution is defined through the osmotic coefficient $\phi(\rho)$ at a  concentration 
$\rho$.
The osmotic coefficient is given through the virial equation~\cite{HansenMcDonald}:
\begin{eqnarray}
\phi(\rho)=1-\frac{\pi}{3}\rho\sum_{i,j} \int_{0}^{\infty} g_{ij}(r;\rho)\frac{d\beta V_{ij}^{\rm eff}(r)}{dr}r^3dr,
\label{eq:phi_virial}
\end{eqnarray}
where the indices $i$ and $j$ represent the two salt components and $g_{ij}(r;\rho)$ needs to be evaluated at the respective
 concentration. 

The virial route, as implemented in this work, is not exact as it employs the infinite dilution pair potential $V_{ij}^{\rm eff}$. 
Accordingly, many-body contributions to the ion-ion interactions for higher densities as induced by the water are not included.
It has been suggested, though, that many-body contributions to the pair-potential can be qualitatively included by taking into account the 
concentration  dependence of the water dielectric constant $\varepsilon(\rho)$ \cite{hess:prl:06,hess:jcp:06}. 
Thus, the long-range part in the pair potential $V_{ij}^{\rm eff}(r)$ has to be altered by using $\varepsilon(\rho)$ instead of the infinite dilution limit $\varepsilon(0)$. The following correction has been shown to lead to agreement of the virial route with the exact compressibility route up to a concentration of roughly $\simeq 2$~M \cite{kalcher09}:
 \begin{equation}
\tilde{V}_{ij}^{\rm eff}(r) =V_{ij}^{\rm eff}(r) -\frac{z_i z_j}{r}[\lambda_B(0)-\lambda_B(\rho)]
 \label{eq:pmf_corr}
 \end{equation}
where again $\lambda_B(\rho)=\beta e^2/4\pi \varepsilon_0 \varepsilon(\rho)$ is the Bjerrum length in the aqueous electrolyte solution of concentration $\rho$, and 
$\varepsilon(0)\sim 72\pm 1$  for SPC/E water \cite{kalcher09}, consistent with previous studies \cite{kusalik94}.
The input parameter $\varepsilon(\rho)$ is directly calculated from the MD simulations and fitted through the function 
 \begin{equation}
\varepsilon(\rho)=\frac{\varepsilon(0)}{(1+A\rho)}
 \label{eq:eps}
 \end{equation}
 where the values of the constant $A$ for each salt are shown in Table \ref{tab:epsA}.
 
 \begin{table}[h]
\begin{tabular}{|c|c|c|c|c|c|c|c|c|c|c|}\hline\hline
& CsCl & KCl & NaI & KF & CsI \\ \hline
A [1/M] & 0.23 & 0.24& 0.34& 0.19&  0.26\\ \hline
\end{tabular}
\caption{Values for the A parameter in Eq. \ref{eq:eps} as calculated from this and previous work \cite{kalcher09}.}
\label{tab:epsA}
\end{table}

\subsection{The hypernetted chain (HNC) approach}

A more efficient yet more approximate 
evaluation of the variation of the osmotic coefficient over a wide concentration range is possible
with integral equation theory based on the  HNC closure~\cite{friedman:71}.  The latter relates the pair correlation function $g_{ij} (r)$ between two particles $i$ and $j$ to the pair potential in an approximative way~\cite{HansenMcDonald}, through:
\begin{equation}
g_{ij}(r)=\exp\Big[ -\beta \tilde{V}_{ij}^{eff} +h_{ij}(r)-c_{ij}(r) \Big]
\label{eq:hnc}
\end{equation}
where $c_{ij}(r)$ is the direct correlation function, and $h_{ij}(r)=g_{ij}(r)-1$ is the total correlation function. 
The HNC equation is closed by the  Ornstein-Zernicke (OZ) equation of liquid state theory 
which relates $h_{ij}(r)$ and $c_{ij}(r)$ \cite{ornstein14}. 
 For an N-component mixture with particle number densities $\rho_n$ the OZ equation is given as
  $h_{ij}(r)=c_{ij}(r)+\sum_{k=1}^{N} \rho_k \int d\vec{r}' c_{ik}(\vec{r}-\vec{r}' )h_{jk}(r')$.
The HNC approach uses the $\varepsilon(\rho)$-dependent  effective pair potential  (Eq. \ref{eq:pmf_corr}) from the MD simulations as input.   Output is the liquid structure in form of the radial distribution functions $g_{ij}(r)$. The osmotic coefficient of the salt solutions under consideration can then be calculated by the virial equation (Eq. \ref{eq:phi_virial}).  This approach has been used before and details can be found elsewhere~\cite{vrbka09}. The derived osmotic coefficient-concentration curves are in good agreement with the MD-derived ones but start to show significant deviations above 1.5-2M.

We note that experiments as well as atomistic MD simulations treat the system at the so-called Lewis-Randall (LR) level while the implicit HNC theory uses the McMillan-Mayer (MM) level \cite{mcmillan}.
At the MM level, the thermodynamic functions are calculated at constant chemical potential of the solvent, thus the MM and LR approaches correspond to different statistical ensembles. The pressure is given by the total pressure of the solution within LR and by the osmotic pressure of the solutes in equilibrium with the solution within MM. In order to compare the results between MD and the HNC approach used here, we follow the conversion:
\begin{equation}
\phi_{MM}=\phi_{LR}(1+mM_s)\frac{\rho_0}{\rho'}=\phi_{LR}\frac{m\rho_0}{\rho},
\label{eq:phi_conv}
\end{equation}
where $m$ and $\rho$ are the molality and molarity of the solution, $M_s$ the molar mass of the solute, $\rho_0$ and $\rho'$ the mass densities of the pure solvent and the solution, respectively \cite{vrbka09}. 
We thus use Eq. (\ref{eq:phi_conv}) to convert the HNC results to the LR level.
Throughout the paper the osmotic coefficients shown are those corresponding to the LR level,
and we use the notation $\phi$ without a subindex.

\subsection{Choice of parameters and optimization procedure}

\begin{center}
\begin{figure}
\includegraphics[width=1\columnwidth]{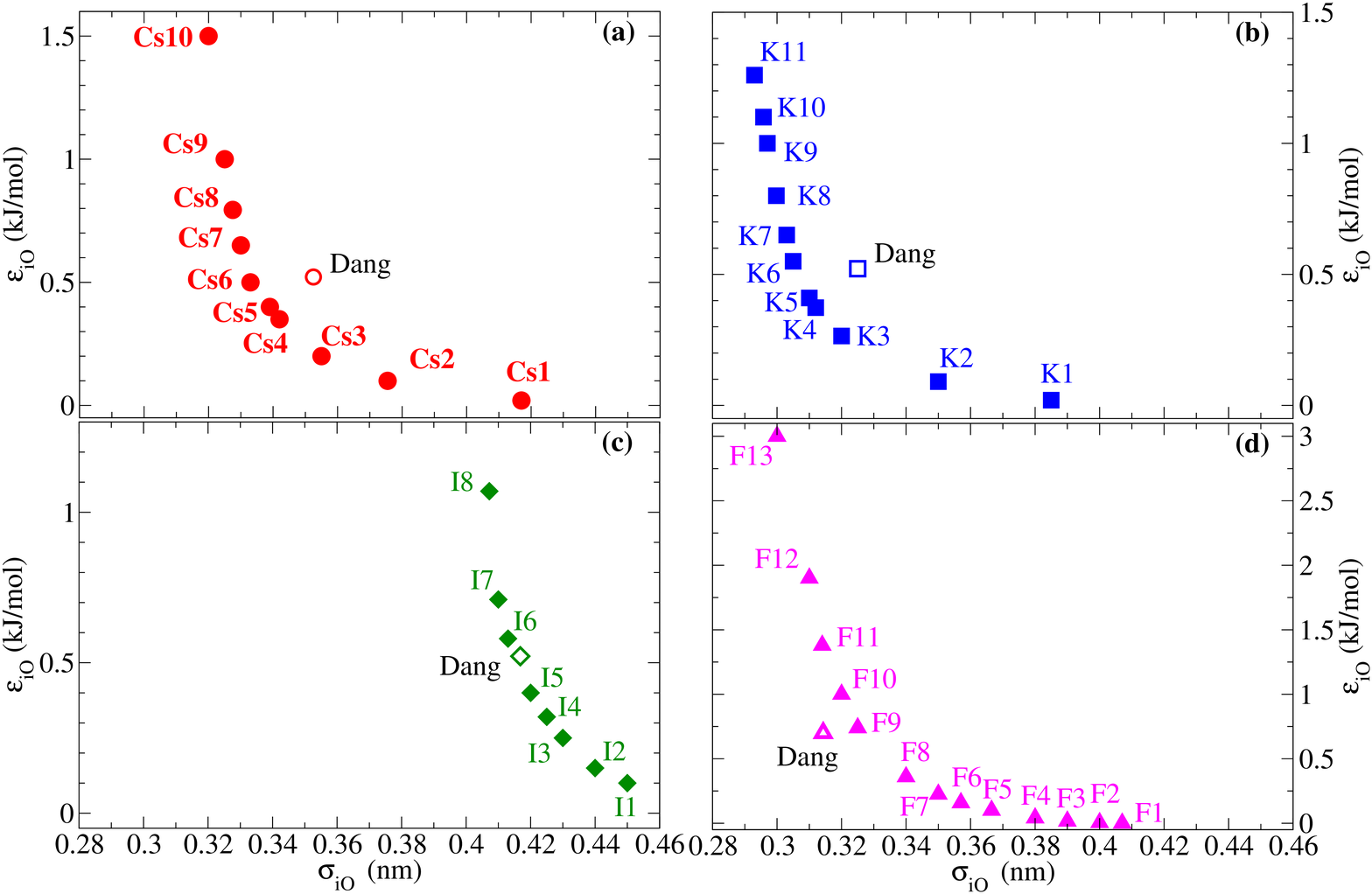} 
\caption{All Lennard-Jones $\varepsilon_{iO}$ and $\sigma_{iO}$ parameters for the ions
(a) Cs$^+$, (b) K$^+$, (c) I$^-$, and (d) F$^-$ used in the optimization procedure in this work. 
All parameters lie on the curves on which the experimental  single ion solvation free energies are 
reproduced~\cite{horinek09}. 
The open symbols are the respective LJ parameters from Dang \cite{dang:jacs:95,dang:92}.}
\label{fig:ion_param}
\end{figure}
\end{center}

As a crucial ingredient to our strategy,
all Lennard-Jones  parameters investigated by us lie on the curve that reproduces the 
experimental free energy of solvation  for a given single ion, which has been calculated previously~\cite{horinek09}.  
This way, we can check whether parameter combinations exist that simultaneously reproduce 
single ion solvation as well as  collective solution properties.  
All Lennard-Jones ionic parameters employed  in this work are 
depicted in Fig. \ref{fig:ion_param}.

The procedure we have followed here for the optimization of ionic force fields is the following: 
for each salt solution and each parameter set used, (a) we start with MD simulations of about 150~ns at a 
concentration of 0.3~M and (b) we simulate the same system for about 30-50~ns at a concentration of 1~M. 
We determine the rdfs between the ions and extract the effective pair potentials from the low-concentration
simulation according to the methodology outlined in~\ref{pp}. 
The effective pair potential and the rdf leads to the osmotic coefficient for the specific solution at 1~M according to the virial route and Eq.~(\ref{eq:phi_virial}). Note that for step (a) the simulation time is longer as more statistics need to be gathered for the computation of the pmfs at 0.3~M, compared to  the rdfs at 1~M concentration. 
In order to obtain the variation of the osmotic coefficient $\phi(\rho)$ for a whole
concentration range, we follow the HNC approach outlined  above. 

Optimization of the LJ parameter for each ion is attempted by comparing the osmotic coefficient at 1~M as calculated from the MD simulations  to the corresponding experimental values. 
The $\phi(\rho)$ curves from the HNC calculations only serve as an indication whether the overall behavior of $\phi$ is reasonably compared to the experimental curves and is not used for parameter optimization.

The salts that were modeled in this work are CsCl, KCl, NaI, KF,
with the goal to obtain optimized force fields for Cs$^+$, K$^+$, I$^-$, and F$^-$.
As a check on the transferability of the obtained parameters, 
we also considered a solution of CsI with parameters optimized for CsCl and NaI. 
NaCl from Dang in SPC/E 
has been found to describe the osmotic properties and activity well when compared to experiments~\cite{kalcher09}, while the individual ions also yield reasonable values for the solvation free energies~\cite{suk:jpcb,horinek09}.  
For this reason, we do not attempt to optimize the force fields of Na$^{+}$ and Cl$^{-}$ and rather 
consider them as a given reference.  
We begin with MD simulations of CsCl, KCl, (for which the Cl$^-$ force field is that from Ref.  \cite{dang:92}) and 
NaI (for which the Na$^+$ force field is that from Ref. \cite{dang:jacs:95}).
MD simulations are performed for those three salt solutions for all ionic parameters of
Cs$^+$, K$^+$, and I$^-$ summarized in Fig. \ref{fig:ion_param}.
The optimal ionic parameters for Cs$^+$, K$^+$, and I$^-$ are then estimated from the comparison of the 
resulting osmotic coefficients to the experimental values. 
At the next step, we use the optimal LJ parameters for K$^+$ 
in simulations of KF solutions and pursue a similar parameter space survey for F$^-$ with the goal
of finding an optimal parameter set for F$^-$.
The choice of the salts NaI and KF is mainly motivated by the fact that the standard force-fields for  those salts gave very poor description of the osmotic coefficients in previous investigations \cite{kalcher09}.
As a consistency check, we finally take the optimized LJ parameters for Cs$^+$ and I$^-$ and 
consider  CsI solutions and compare the  calculated osmotic coefficient 
$\phi$ for CsI  to experiments.

A well-known problem of ionic force fields \cite{auffinger07} has to be mentioned. Our MD simulations show that very low $\varepsilon_{iO}$ parameters for the cations (Cs$^+$ and K$^+$) lead to unphysical  clustering of the ions for CsCl and KCl even at low concentrations of 0.3~M. Accordingly, no pmfs could be extracted and the corresponding parameters (Cs1, Cs2, K1, K2, K7, and K9) are neglected for the calculation of the osmotic coefficient and for the optimization procedure of the Cs$^+$ and K$^+$ force fields. On the other hand, very low $\varepsilon_{iO}$ for the anions (I$^-$ and F$^-$) do not lead to similar clustering for NaI and KF, respectively, and can still be considered as candidates for an optimized force field. Interestingly, ion clustering was observed for all salts at a concentration 1M in some regions of the LJ parameter space. Specifically, the ions in KF clustered for the parameter sets F12, F13, for which the K11 parameter was used.  For F1, F2, we could not obtain reasonable results, as the ion-water system could not be energetically optimized within MD, thus those parameters had to be eliminated as well. Apart from these restrictions, we have used all other parameters in Fig. \ref{fig:ion_param} 
for the MD simulations. Parameter sets for which results are not shown for the osmotic coefficients in Fig. \ref{fig:phis1M} are the ones that led to clustering of the ions in the aqueous environment.

\section{Results and discussion\label{sec:result}}

\subsection{Structural properties}

\subsubsection{Radial distribution functions\label{sec:rdf}}

\begin{figure}
\includegraphics[width=1\columnwidth]{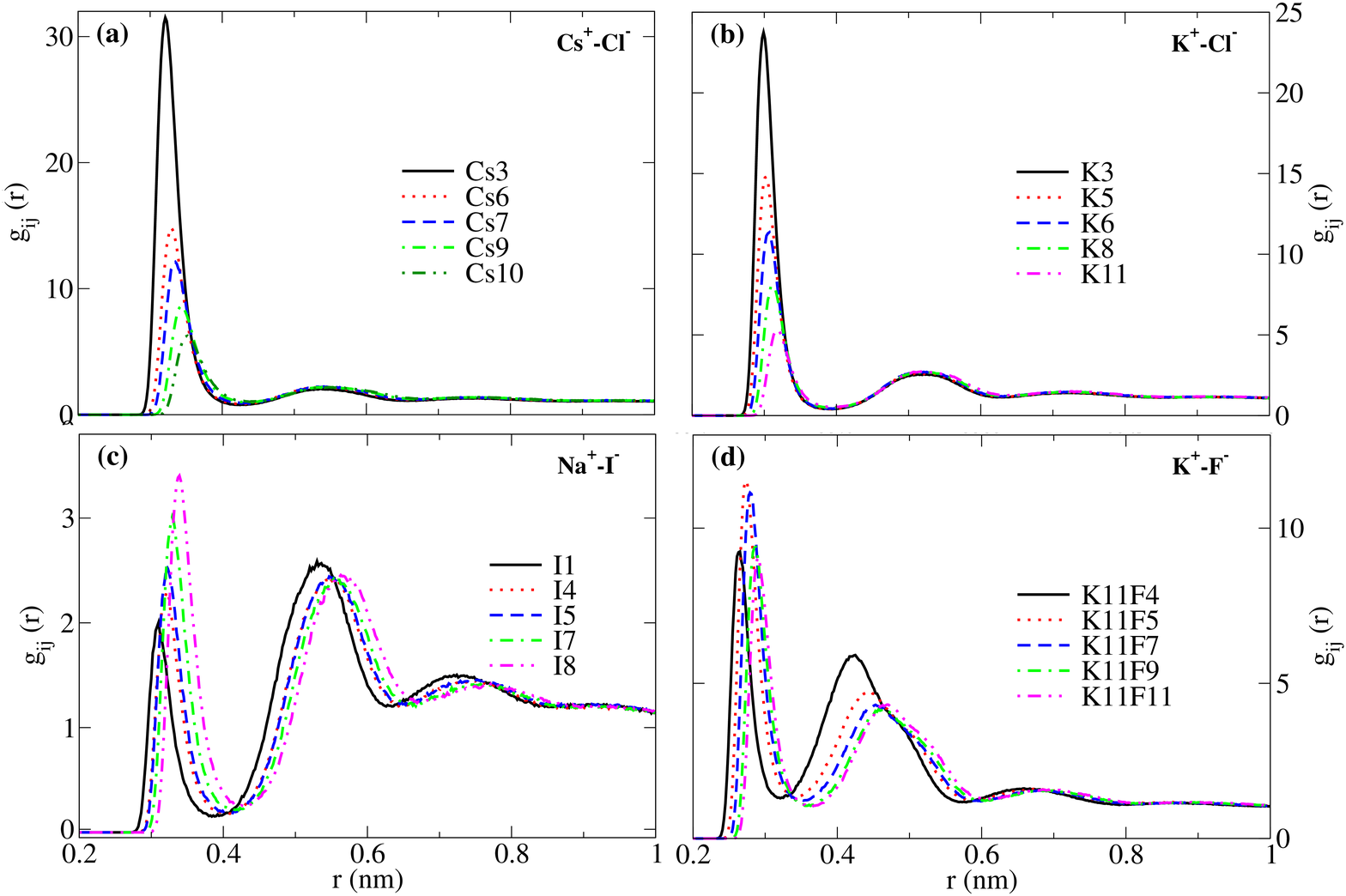}
\caption{Cation-anion radial distribution functions (rdfs) at a concentration of 0.3~M for various parameter sets for (a) CsCl, (b) KCl, (c) NaI, and (d) KF. The Cl$^-$ and Na$^+$ force fields are taken from Dang \cite{dang:jacs:95,dang:92}, the K$^+$ force field in (d) is K11.}
\label{fig:rdfs_param}
\end{figure}

A robust measure of electrolyte structural properties is the ion-ion radial distribution function (rdf), which  shows distinct structural signatures that differ among different salt solutions. Here, we have calculated the rdfs for all systems at two concentrations, 0.3 M and 1 M, respectively. The rdfs for the two concentrations and the same LJ parameters do not differ significantly, apart from the height of the rdf peaks. The heights of the contact and the solvent-separated peak indicate different hydration properties of the ions. All curves exhibit strong electrostatic screening and reach the asymptotic value of 1 below a distance of 2 nm for the 1 M concentration. This is consistent with the small screening length of about 0.25 nm at 1 M. As  shown in Fig. \ref{fig:rdfs_param}, for CsCl and  KCl the first peak in the cation-anion rdfs is much higher than the second one, indicating close contact of the anion and cation and predominant direct ion pairing.  For KF and NaI, the first peak in the cation-anion rdfs is of similar height or  lower than the second one, indicating that water enters between the anion and cation indicative of  indirect ion pairing. Due to the variety of LJ parameters used in this study, the relative strength of direct and indirect ion pairing for the different salts shows rich behavior~\cite{collins1,dill09}. 
The height of the first rdf peak at 0.3~M for  CsCl ranges from about 6 to 30, for KCl from 5 to 25, 
brought about by variations of the cationic force field parameters, as illustrated in Fig. \ref{fig:rdfs_param}.
For the ion pairs that show pronounced solvent-separated pairing, NaI and KF, the influence is much less,
here the height of the first rdf peak varies for  NaI from 2.5 to 3.5 and  for KF from 9 to 12. 
Note that these variations are induced by changing the anionic force fields.
This demonstrates that by changing force field parameters that keep the single-ion properties invariant
(as judged by the ion-water rdf or the  solvation free energy), 
the  ion-pairing properties can be significantly affected \cite{dang:jcp:92}. 
The position of the contact peak also shows interesting behavior. 
 As an example, we present the trends for some representative LJ parameters for the KCl cation-anion rdfs:
for the order of the bare  LJ radius $\sigma_{KO}$, 
$\sigma_{KO}(K{11})<\sigma_{KO}(K{8})< \sigma_{KO}(K{6})< \sigma_{KO}(K{5})<\sigma_{KO}(K{3})$,
 the ordering of the contact peak position between K$^+$ and Cl$^-$,  r$^{cp}_{KCl}$ is 
 $r^{cp}_{KCl}(K{3})<r^{cp}_{KCl}(K{5})<r^{cp}_{KCl}(K{6})<r^{cp}_{KCl}(K{8})<r^{cp}_{KCl}(K{11})$,
 as is visible from Fig. \ref{fig:rdfs_param}.
We find similar trends for CsCl, NaI, and KF, as well. 
Ions with smaller bare Lennard Jones radius thus show larger ion-ion separation, which clearly has to do with the fact
that  the interaction strengths ($\varepsilon_{iO}$) of the single ions are  varied along with $\sigma_{iO}$.
In addition to the peaks of the rdfs we also study the minima in the rdfs. 
The positions of the first and second minima in the rdfs, $r_1$ and $r_2$, and the distance at which the rdfs vanish, $r_0$,
are given for representative salt parameters at a concentration of 0.3~M in Table \ref{tab:rdf}. 

\begin{table}[h!]
\begin{tabular}{|c||c|c|c|c|c|c|c|}\hline
& Cs$6$Cl&Cs$9$Cl&K$5$Cl&K${11}$Cl&NaI$1$&NaI$4$&K11F$5$ \\ \hline \hline
r$_0$ (nm)&0.29&0.30&0.27&0.28&0.28&0.28&0.29 \\ \hline
r$_1$ (nm)&0.43&0.43&0.40&0.40&0.41& 0.40&0.43 \\
r$_2$ (nm)&0.67&0.68&0.63&0.63&0.64&0.65& 0.66\\ \hline \hline
\end{tabular}
\caption{The rdf characteristics at 0.3~M for the optimized ion parameters; see Tab.~III. 
The distance (r$_0$) at which the rdf vanishes to zero and that of the first (r$_1$) and second (r$_2$) minimum in the cation-anion rdfs for representative parameters of Cs$^{+}$, K$^{+}$, I$^{-}$, and F$^{-}$ in CsCl, KCl, NaI, and KF aqueous environments, respectively.}
\label{tab:rdf}
\end{table}

\begin{figure}
\includegraphics[width=0.8\columnwidth]{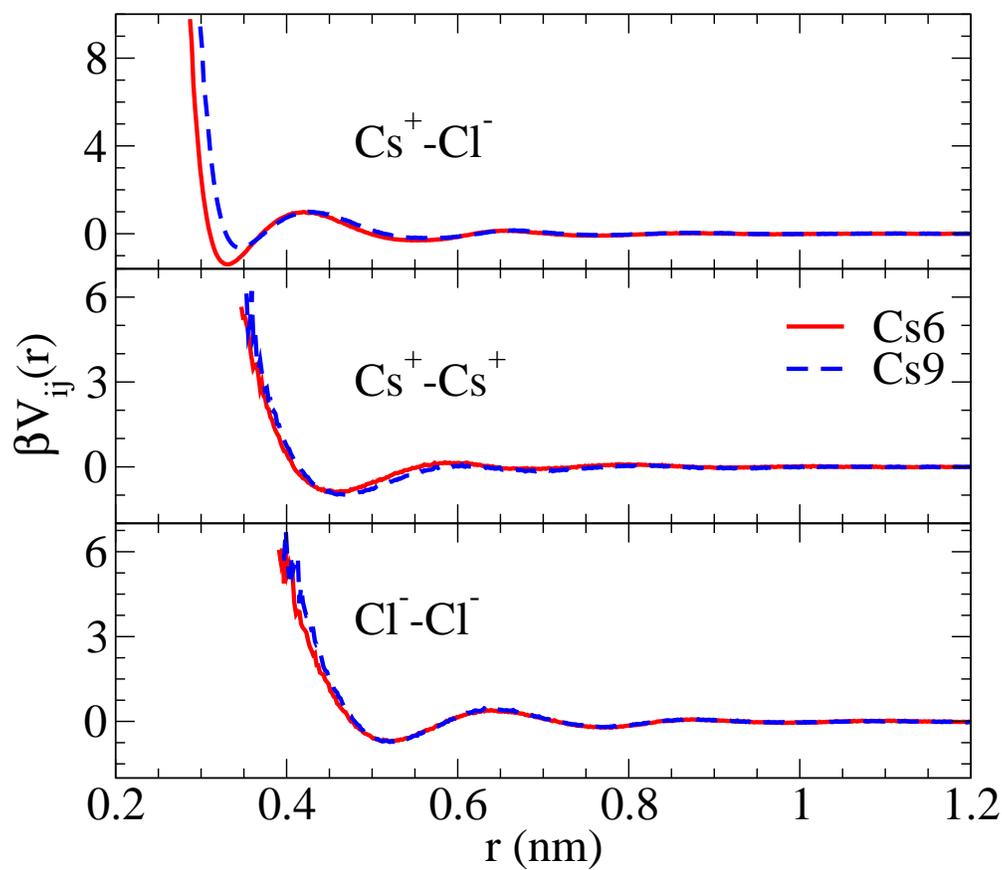}
\caption{Radial distribution functions (rdfs) at a concentration of 0.3~M for the Cs$6$ parameter set. Red (solid) lines correspond to the MD results and blue (dashed) lines to the rdfs as generated through the HNC approach (based on the pmfs from MD). From top to the bottom, the rdfs for the cation-anion, cation-cation, and anion-anion are shown.
}
\label{fig:rdfs}
\end{figure}

We have also compared the $g(r)$s from our MD simulations with results from
 HNC for a few different  salt parameters and observed only small differences; see Fig. \ref{fig:rdfs}.
 The reason for the deviations is the approximate treatment of statistical mechanics in HNC. 
 The differences are mostly visible
 in the height of the first and second peaks in the $g(r)$.

\subsubsection{Short-ranged potentials of mean force}

\begin{figure}
\includegraphics[width=1\columnwidth]{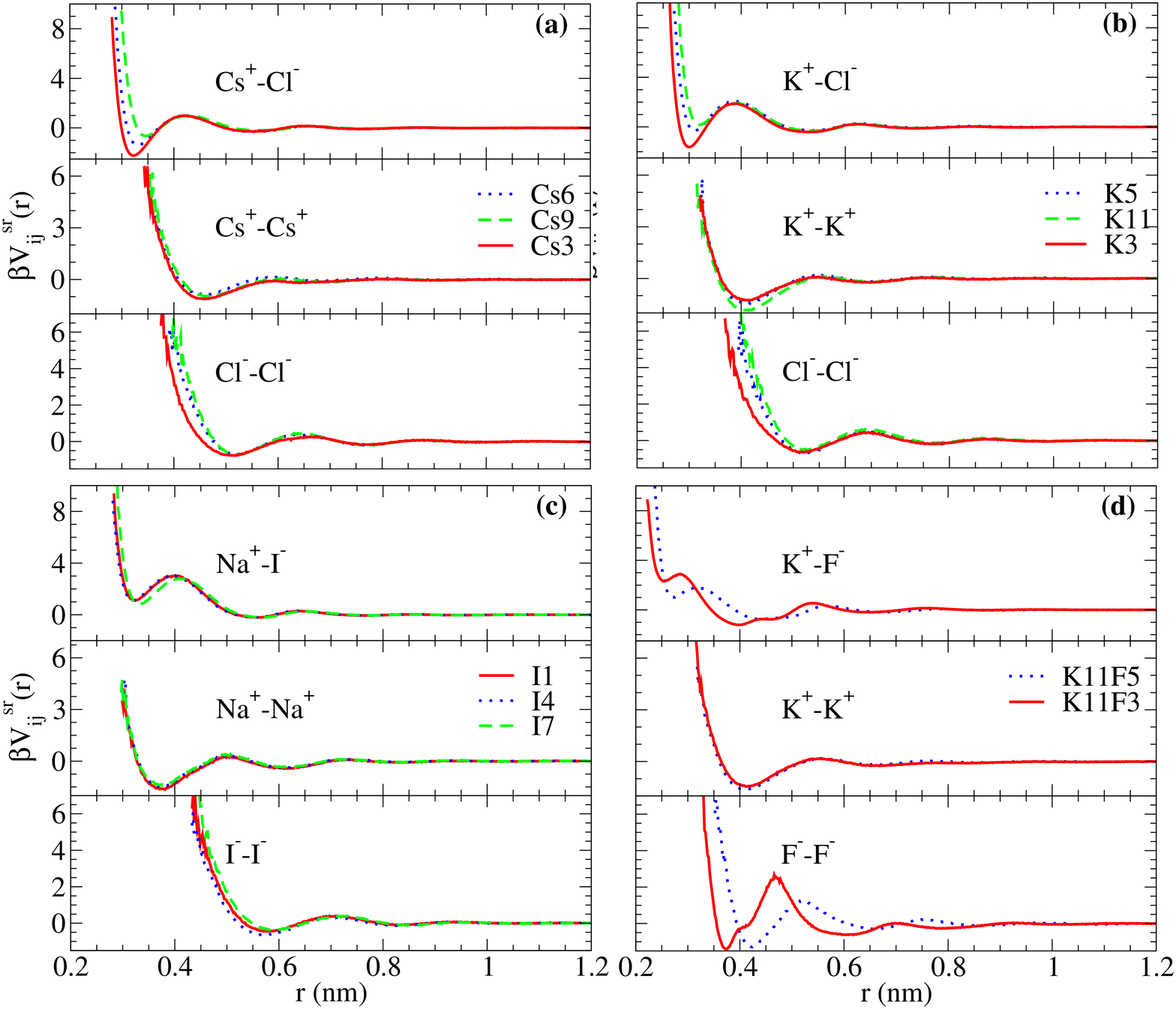}
\caption{Short-ranged potentials of mean force (pmfs) at a concentration of 0.3~M for the optimal   LJ parameters 
(see Section \ref{sec:optim}) for (a) Cs (sets Cs$6$, Cs$9$), 
(b) K (sets K$5$, K${11}$), (c) I (sets I$1$, I$4$) and (d) KF (set K11F5). 
In each panel, from top to the bottom, the pmfs for the cation-anion, cation-cation, and anion-anion are shown. For comparison, in all panels the pmfs for a non-optimal parameter set is included: (a) set Cs3, (b) K3, (c) I7, and (d) K11F3.
}
\label{fig:pmfs}
\end{figure}

We derive short-ranged pair potentials for all salt parameters that do not lead to ion clustering  
from Eq.~\ref{eq:vsr}. Examples are shown in Fig.~\ref{fig:pmfs}. 
In accordance with the rdf's, the pair potentials for NaI and KF 
reveal deeper second minima, while the first two minima for CsCl and KCl show roughly similar depth. 
The second  minimum of KF is broader than for the other salts, 
indicating an unusual water configuration in the solvent-separated ion pair.
The Cs$^+$-Cs$^+$ and K$^+$-K$^+$ pair potentials show smaller oscillations compared to the Na$^+$-Na$^+$ 
potential. This has been observed before, and was rationalized  by the fact that Na$^+$ has tightly bound 
hydration shells giving rise to energy barriers when two sodium cations approach \cite{kalcher09}. 
In the case of the anion-anion pmfs, the F$^-$-F$^-$ pmf shows deeper minima and stronger oscillations
than the  Cl$^-$-Cl$^-$ or the I$^-$-I$^-$ potential. 
Going from the small fluoride to the large iodide, one observes a trend towards a soft-sphere like potential. 
Similar to the rdfs, we do not observe a simple dependence of  the position and depth of the minima in the pmfs with the variation of the LJ parameters, again due to the simultaneous change of both the LJ radius and interaction strength
along the lines of constant solvation free energies.
Note again that there is substantial variation in the cation-anion potentials for the different force-field parameters,
which gives hope to be able to fit the osmotic coefficients. The scattering in the Cl$^- -$Cl$^-$
potentials (note that the Cl$^-$ force field is not changed) is due to bad sampling statistics as the 
ions typically do not get very close.

\subsection{Osmotic coefficients}

\begin{figure}
\includegraphics[width=0.6\columnwidth]{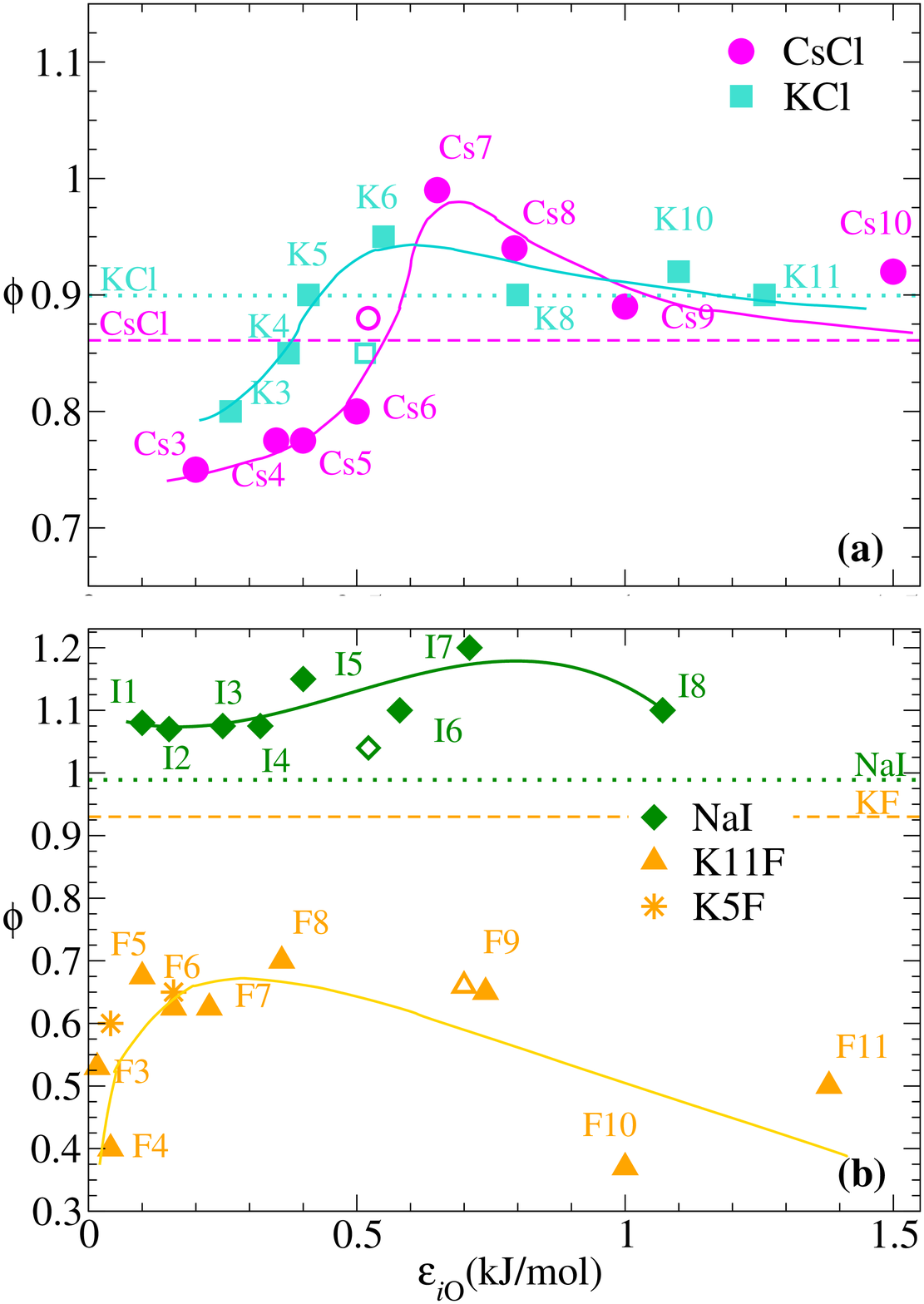}
\caption{The osmotic coefficient $\phi$ as calculated from the MD simulations by means of the virial route for CsCl, KCl, NaI and KF in a 1~M solution. In (a) the magenta(dashed) and turquoise(dotted) lines represent the experimental results for CsCl and KCl, respectively. In (b) the green (dotted) and orange(dashed) lines are the experimental values for NaI and KF, respectively. Representative results for KF for the K5 parameter set are also shown. The open symbols are the 
osmotic coefficients for the Dang parameters obtained in a  previous work \cite{dang:92,dang:jacs:95,kalcher09}. The solid lines are guides to the eye.}
\label{fig:phis1M}
\end{figure}

Having determined the rdfs and short-ranged pair potentials for different LJ parameters, 
we next calculate the osmotic coefficient $\phi$  using the virial route based on
the rdfs and pair potentials from MD simulations (not HNC) as explained in Section \ref{sec:method}. 
The results for CsCl, KCl, NaI, and KF at a concentration of 1~M are summarized in Fig. \ref{fig:phis1M}
where we have added lines as guides to the eye in order to bring out the main features
of the results more clearly.
The error bars for the calculated $\phi$ are in the range $\pm$0.01-0.05.
The experimental values of the osmotic coefficients \cite{heydweiller10,hamer72} 
for each salt are also shown, on which we  base our optimization of the ionic force fields. 
Inspection of this figure reveals that for the  salts CsCl, KCl, and KF  $\phi$ 
shows a maximum for intermediate values of $\varepsilon_{iO}$.
For NaI,  no significant variation of $\phi$ with $\varepsilon_{iO}$ is observed. 
We first focus on the cations Cs$^+$ and K$^+$ in CsCl and KCl, respectively, and panel (a) in Fig. \ref{fig:phis1M}. 
As more clearly shown by the lines added as guides to the eye  for the CsCl and KCl data, 
there are two crossings between the simulated and the experimental values for $\phi$,
so in principle there are two optimal force field sets for these cations.
Note that we also included $\phi$ for the Dang parameters for KCl and CsCl~\cite{dang:92,kalcher09} as open symbols.
Interestingly, though they are somewhat off from the curve corresponding to the experimental solvation free energy
(see Fig. \ref{fig:ion_param}), they show quite good agreement with the experimental data for $\phi$  within the error.  

Inspection of panel (b) reveals that the situation for the anions I$^-$ and F$^-$ is distinctively different. 
For all iodide parameters, the corresponding $\phi$ values are close to the experimental value (within the error), 
but show very little variation. 
This suggests that varying the force field for I$^-$ has no considerable effect on the osmotic coefficients. 
A very similar result was obtained for chloride in NaCl in previous simulations~\cite{smith:jcp:03}.
However, this does not seem to be generally true for anions, as 
KF in the same panel shows a much larger variation in $\phi$, ranging from about 0.4 up to 0.7. 
For most of the results for KF shown in Fig.~\ref{fig:phis1M}(b), 
the optimized K$11$ parameter set from Table \ref{tab:optim} was used. 
However, as seen in the figure,  the use of the equally optimal  force field  K$5$ does not lead to qualitative changes.
For all F$^-$ parameters, the osmotic coefficients of KF are too low compared to  the experimental osmotic coefficient
and even for the best F$^-$ parameter combination are too low by about 0.25, which is larger than the error bars.
We conclude that while force fields for Cs$^+$ and K$^+$ can be derived with relative ease, the situation is more
complicated for the anions considered by us: while I$^-$ works quite well (which seems to be coincidental since
the variation of $\phi$ with $\varepsilon_{IO}$ is very small), the optimization for F$^-$ fails though here the variation of $\phi$ with $\varepsilon_{IO}$ is pronounced.

\begin{figure}
\includegraphics[width=1\columnwidth]{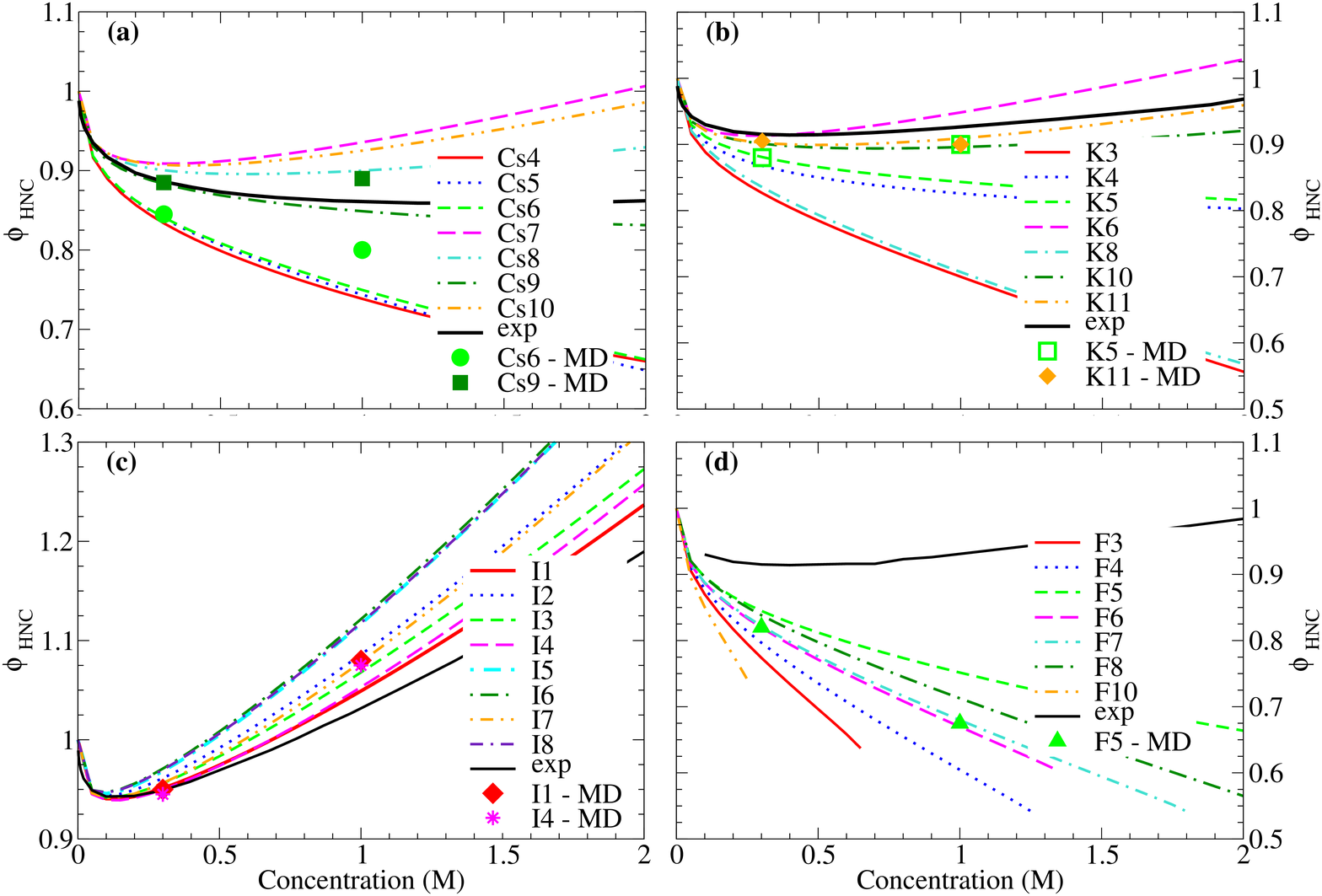}
\caption{The osmotic coefficient $\phi$ as calculated through the HNC approach (lines, see text) of representative 
LJ parameter sets for (a) CsCl, (b) KCl, (c) NaI, and (d) KF as a function of concentration (M). The experimental curves \cite{heydweiller10,hamer72} are also shown. The symbols are the values directly from MD for the parameters sets proposed in Table \ref{tab:optim}. In panel (d) all KF results are obtained by using the K11 parameter set for the potassium ion. 
}
\label{fig:phisHNC}
\end{figure}

Using the HNC approach, we have also calculated the variation of the osmotic coefficient with the concentration for all salts and LJ parameters investigated here. Curves for some of the parameter sets used are shown in Fig. \ref{fig:phisHNC} together with experimental data. For some of the LJ parameters for fluoride, $\phi$ for KF diverges above a certain concentration, for those cases the corresponding curves are cut above the concentration of 2M. Inspection of the overall shape of the curves reveals that there is good agreement with the experimental curves for some LJ parameters for CsCl, KCl and NaI, in contrast to KF.  This is not unexpected, since for KF simulation results for $\phi$ did not match the experimental value at 1 M for any parameter combination. Note that there are small differences between the HNC results (lines) and MD results at 0.3M and 1m (symbols)
for $\phi$,  which are of the order of the error inflicted by the virial approximation 
and the simulation numerical scatter of about $\pm$ 0.05, see our previous discussion \cite{vrbka09}.

\subsection{Optimum ionic force field parameters\label{sec:optim}}

Based on the MD osmotic coefficient results, we suggest optimal LJ parameters for 
Cs$^+$, K$^+$ for use with Cl$^{-}$ parameters from Dang \cite{dang:92}, 
and optimal parameters for I$^-$ for use with Na$^{+}$ from Dang \cite{dang:jacs:95};
we also suggest a force field for F$^-$ derived from simulations of KF using our optimized force field for K$^+$. For most ions, two parameter sets are suggested and summarized in Table \ref{tab:optim}. For the cations, Cs$^+$ and K$^+$, we take two values closest to the crossing points of the fitting curves for $\phi$ with the experimental data, see Fig.~\ref{fig:phis1M}, namely Cs$6$, Cs$9$, K$5$, and K$11$. 
For K$^+$ a slight ambiguity exists, as for example also  the K$8$ force field matches the experimental data. 
We rather chose  K$11$ because that parameter is further away from the parameter sets K$7$ and K$9$ for which the ions were found to cluster. For iodide, our choice of parameters is less obvious; basically all parameters within the error bars are equally acceptable.

\begin{table}[h!]
\begin{tabular}{|c||c|c|}\hline\hline
ion/label &$\sigma_{iO}$ (nm) &$\varepsilon_{iO}$(kJ/mol) \\ \hline
Cs$6$ & 0.333& 0.5 \\ \hline
Cs$9$ & 0.325& 1.0 \\ \hline \hline
K$5$ & 0.31& 0.41\\ \hline
K${11}$ & 0.293& 1.26\\ \hline \hline
I$1$ & 0.45& 0.1\\ \hline
I$4$ & 0.425& 0.32\\ \hline 
F$5$ & 0.3665& 0.1\\ \hline\hline 
\end{tabular}
\caption{The optimized Lennard-Jones parameters for Cs$^+$, K$^+$, I$^-$, and F$^-$ as extracted by the MD simulations and the comparison of the osmotic coefficients with experimental data.}

\label{tab:optim}
\end{table}

For KF, we performed two distinct sets of MD simulations, the first with the K$11$ force field and 
the second with the K$5$ parameters. Interestingly, both K$^+$ give comparable results for $\phi$, see Fig.~\ref{fig:phis1M}b, meaning that the redundancy found with KCl seems to be also present for KF. However, none of the KF parameters reproduces experimental values for $\phi$, so we chose a single force field for F$^-$ that shows the least deviation.

In Fig.~\ref{fig:pmfs}, pmfs for  the optimal parameters are compared with a pmf of a  non-optimal LJ parameter set. As expected, for Cs$^+$ and K$^+$ the cation-anion pmf of the non-optimal force field shows larger deviations from the optimal force field results; for I$^-$ all anion-cation pmfs are quite similar. 

We briefly return to the discussion on peak heights and ion pairing~\cite{collins1,dill09}. For the optimal force fields in Table~\ref{tab:optim}, we find for the height of the first contact peak in the rdfs at 0.3~M concentration, values of about 9.5 and 5.5 for K$5$Cl, and K${11}$Cl, respectively; 14.8 and 8.6 for Cs$6$Cl and Cs$9$Cl, respectively; about 2.3 for both NaI$1$ and  NaI$4$, respectively, and 3.3 for K$11$F$5$.
The order of the peak height is KCl $\sim$ CsCl $>$ KF $>$ NaI, 
consistent with previous theories on  ion pairing \cite{collins1,dill09}, according
to which the tendency to form direct ion pairs goes down as the ion sizes become more dissimilar.
We note that this ordering of contact pair formation probability is only realized for the optimized 
force fields, non optimal force field combinations can easily lead to partial or complete reversal of this ordering.

\subsection{Transferability of the optimized ionic force fields}

\begin{figure}
\includegraphics[width=1\columnwidth]{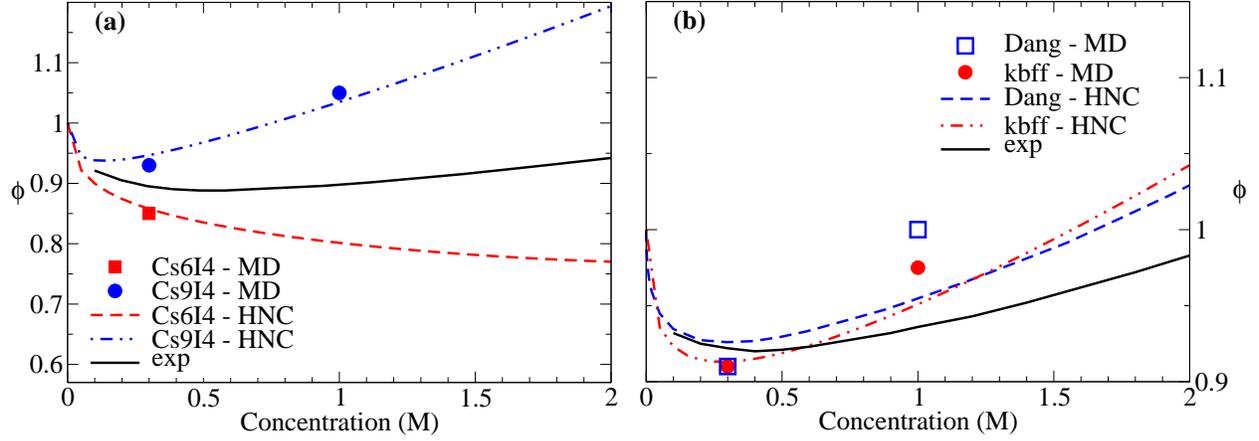}
\caption{MD results (symbols) for the osmotic coefficient, plotted together with the HNC results (broken lines)
as a function of  the salt concentration (in M). Panel (a) shows results for CsI and serves as a check on the transferability 
of the proposed force fields for Cs$^{+}$ and I$^{-}$. Panel (b) depicts the results for NaCl from the Kirkwood-Buff derived force field
(kbff) ~\cite{smith:jcp:03} and the Dang parameters~\cite{dang:jacs:95,dang:92,kalcher09}.  In both panels, the black (solid) lines correspond to the respective experimental curves \cite{heydweiller10,hamer72}.}
\label{fig:phisCsI_NaCl}
\end{figure}

We so far were occupied with finding force field parameters that match experimental osmotic coefficients best.
We now turn to a separate question and 
check whether the optimized force fields presented in the previous section 
are transferable. To that end, we perform a set of MD simulations for CsI in water, for which the parameters 
for Cs$^+$ and I$^-$ are taken to be the optimized force fields from Table \ref{tab:optim}. 
The hope can be not be to match experimental osmotic coefficients for CsI perfectly, as the Iodide
parameters are not perfect by themselves (when compared with experimental osmotic coefficient data for NaI).
Rather, we intend to check whether the trends of the simulated osmotic coefficients of CsI reflect the 
properties of the Cs$^+$ and I$^-$ force fields. 
The specific salts modeled are Cs6I4 and Cs9I4 (we did not check the I$1$ parameter set, as
variations of iodide parameters do not seem to considerably affect the osmotic coefficients, as was seen
from Fig.~\ref{fig:phis1M} for NaI). The MD and HNC results for the osmotic coefficient are 
shown in Fig. \ref{fig:phisCsI_NaCl}(a).  Note that for Cs6I4 we only show the MD data for 0.3M,
since the simulation data at 1M show bad convergence behavior, probably due to the vicinity
to the crystalization transition for this particular force field (note that experimentally CsI has the smallest maximal 
solubility of all considered salts in this paper, of about 3M, so such problems might be anticipated).
We find  overall good agreement between the MD and HNC results with deviations in 
$\phi$ smaller than 0.05. There are quite sizeable deviations between the 
theoretical predictions and the experimental data. However,
note that we have used the optimized parameters for both Cs$^+$ and I$^-$,
based on osmotic coefficients for CsCl and NaI, and that  
the deviations from experimental data for CsI in Fig. \ref{fig:phisCsI_NaCl}(a) are
of the same order as the deviations observed for 
CsCl and NaI in Fig.~\ref{fig:phis1M}.
Furthermore, the deviations from experimental data go in the expected direction, namely, the 
simulation prediction for Cs6I4 lies below the experimental value, while Cs9I4 lies above (similar to the data
for Cs6Cl and Cs9Cl in Fig.~\ref{fig:phis1M}a). 
We tentatively conclude from these data that the force fields obtained by our optimization strategy are transferable,
meaning that if one fixes the force fields of ions A$^+$ and D$^-$ by   optimizing the 
osmotic coefficients of the salts  AB and CD, the osmotic coefficient of the salt AD should come out 
approximately right without further adjustment. Such reasoning 
of course assumes that one can optimize the salts AB and CD and therefore does 
not apply to F$^-$, where the optimization of KF fails in the first place. So one sees that while optimization and 
transferability are logically distinct operations, transferability presumes successful optimization of force field parameters.
We will come back to this point in the conclusions.

As an additional check of our methodology, we also used a different force field for NaCl that was
designed to reproduce experimentally determined Kirkwood-Buff integrals
(and thus experimental activity coefficients) \cite{smith:jcp:03}.  
We performed simulations with those  NaCl parameters using the mixing rules proposed, 
namely geometric mixing including a freely adjusted mixing coefficient 
(and not the Lorentz-Berthelot rule which we were using up to this point). 
The MD and HNC results for the osmotic coefficient are shown in Fig.~\ref{fig:phisCsI_NaCl}(b), 
together with the experimental data and those derived previously 
from the NaCl Dang parameters~\cite{dang:jacs:95,dang:92,kalcher09}.  
It is evident from this figure that the Kirkwood-Buff NaCl force fields~\cite{smith:jcp:03}
 give good agreement with experimental values for the osmotic coefficient,
 indicating that activity coefficient and osmotic coefficient data probe the same ionic features, namely
 the pairing characteristics in aqueous solution.

\section{Summary and conclusions\label{sec:concl}}

We have used a combination of MD simulations and statistical mechanics analysis 
to systematically explore and optimize force field parameters for ions in aqueous solution,
with particular emphasis on the interplay of single-ion and ion-pair thermodynamics.

The ion parameters considered by us, namely the Lennard-Jones radius and strength, 
were confined to a curve on which the experimental  single ion solvation free energy  is reproduced~\cite{horinek09}. For a whole number of  specific force fields on those curves,
we constructed effective ionic pair potentials  which led, through the virial route, to electrolyte osmotic coefficients. These were then compared to  experimental values at 1~M concentration.  
We have used the MD-virial route derived osmotic coefficients to optimize the LJ parameters for the single ions Cs$^+$, K$^+$, I$^-$, and F$^-$ based on data for  CsCl, KCl, NaI, and KF, respectively, treating Na$^+$ and Cl$^-$ as reference ions.  This optimization strategy works well for the cations Cs$^+$ and K$^+$, for which we obtain, due to the peculiar shape of the MD-based osmotic coefficient curves, two optimized force fields for each ion. This proves that at least for the cations, the simultaneous description of single-ion and ion-pairing thermodynamics is possible if one systematically explores the full LJ parameter space without the need to modify the combination rules. 
For anions, on the other hand, the optimization is more problematic.
We could not get  reasonable match of  experimental osmotic coefficient data varying  the LJ parameters 
of F$^-$ in  KF, simply because the maximum in the simulated osmotic coefficient is significantly below the experimental value.
Iodide can be more or less satisfactorly  optimized based on NaI data, but this seems to be mere coincidence, since
NaI  osmotic coefficients shows very little dependence on the LJ parameter, 
which indicates a basic limitation of our approach.
With these restrictions in mind, we checked for the transferability of our optimized force fields by 
considering the osmotic coefficient of a CsI solution, i.e. an ion  combination that was not targeted in the 
optimization process. Based on the far-from-perfect performance of the I$^-$ force field in NaI, we judge the transferability
properties of the force fields as satisfactory since in particular the trends of the CsCl osmotic coefficients upon
variations of the Cs$^+$ force field are fully recovered when looking at the trends of the CsI osmotic coefficients.

These results suggest that for truly accurate non-polarizable  ion parameters, one apparently needs to lift the constraint of the simple mixing rules  and introduce an additional scaling parameter in the mixing rule, in agreement with previous approaches \cite{smith:jcp:03,hess09}. However, it seems desirable and possible to introduce such a mixing parameter only in the cation-anion interaction, so that the anion-water and the cation-water interactions stay the same as used in the single-ion solvation free energy optimization (the cation-cation and anion-anion interactions are much less important).  That way, one would correctly describe the single-ion properties (as embodied in the infinite-dilution solvation free energy) as well as ion-pairing properties (as important for osmotic and activity coefficients). As an unfortunate byproduct, one would have to optimize this mixing parameter for each ion pair, requiring substantial simulation efforts.

\acknowledgments

The authors wish to thank D. Horinek for discussions.
We acknowledge support by
the Ministry for Economy and Technology (BMWi) in the framework of the AiF project
 "Simulation and prediction of salt influence on biological systems",
by the NIM Cluster of Excellence in Munich and the Emmy-Noether Program of 
the Deutsche Forschungsgemeinschaft (DFG). MF acknowledges support from the 'Gender Issue Incentive Funds' of the Cluster of Excellence in Munich.
The Leibniz Rechenzentrum Munich is acknowledged for supercomputing access.

\newpage

 \newpage

\newpage

\newpage

\newpage

\newpage

\newpage

\end{document}